# Hexagonal boron nitride crystal growth in the Li$_3$BN$_2$-BN system

Camille Maestre[1,2*], Philippe Steyer[2], Bérangère Toury[1], Catherine Journet[1*], Vincent Garnier[2]

*1. Universite Claude Bernard Lyon 1, CNRS, LMI UMR 5615, Villeurbanne, F-69100, France*

*2. INSA Lyon, Universite Claude Bernard Lyon 1, CNRS, MATEIS, UMR5510, 69621 Villeurbanne, France*

[*]Authors to whom correspondence should be addressed :
camille.maestre@univ-lyon1.fr
catherine.journet@univ-lyon1.fr

## Abstract

Hexagonal boron nitride (hBN) presents valuable intrinsic properties and attracts considerable attention for the development of novel two-dimensional (2D) materials-based technologies. Even though huge efforts have been made to improve the bottom-up synthesis of integrated and high quality hBN, the devices presenting the best performances are still made using hBN exfoliated from bulk crystals. In this context, we explore the Polymer-Derived Ceramics (PDC) route coupled to a high temperature process that produces millimetric and high quality hBN crystals. By investigating the (micro)structure of several samples, we demonstrate that the crystal growth occurs by segregation from a Li$_3$BN$_2$-BN solution upon cooling and from hBN seeds. In particular, we show that crystallization can occur at a temperature as low as 1400°C. Overall, these results show that hBN crystal growth in the Li$_3$BN$_2$-BN system is compatible with conventional flux methods that may be the most promising platform for continuous seeded hBN crystal growth.

## I. Introduction

Hexagonal boron nitride (hBN) is a chemically and thermally stable material that, when exfoliated, forms two-dimensional (2D) nanosheets. The electrically insulating nature of 2D hBN, combined with its atomically flat surface and its high thermal conductivity, makes it a crucial building block for 2D materials-based technologies [1]. 2D materials are indeed very sensitive to the dielectric environment, and defect-free hBN is therefore required to make the most of them [2]. Moreover, thicker and bulk hBN can also be of interest for UV laser [3], heat dissipation [4] or neutron detection [5] applications. Many groups are intensively investigating the hBN thin film deposition with significant improvements [6–8]. Yet, due to their higher crystallinity, hBN nanosheets (BNNS) exfoliated from high-quality freestanding crystals are still preferred for demanding applications [9–11].

Millimeter-sized polycrystalline hBN has initially been obtained as a cubic boron nitride (cBN) side product during its synthesis at high pressure high temperature (HPHT) [12]. The HPHT process is described as a Ba-BN solution growth synthesis driven by a thermal gradient [3,13], yet its scaling is limited by the harsh conditions involved. Another hBN growth process, the Atmospheric Pressure High Temperature (APHT), was demonstrated later using metallic solvents (Fe or Ni based alloys) [14–19]. The crystal growth is usually driven by slow cooling, but the use of a thermal gradient has been reported as well [20]. A full review on self-standing hBN crystal synthesis is available elsewhere [21]. The APHT process is however limited by the low solubility of B and N atoms in the metal flux and by the dissolution kinetics.

Concurrently, a specific chemical approach coupled with sintering techniques has been developed: the polymer-derived ceramics (PDC) route. This process applied to hBN [22–26] involves an intermediate compound polycondensed from borazine ($B_3N_3H_6$), the polyborazilene (PBN, $(B_3N_3H_x)_n$). It has been shown that PBN can be ceramized into BN of various levels of crystallinity at moderate temperatures [27]. In particular, the addition of $Li_3N$ to PBN could favor the conversion of PBN into crystallized hBN [25,27,28]. The ceramization of PBN mixed with $Li_3N$ in a pressure controlled furnace (PCF) has recently given rise to large hBN crystals [29,30] presenting excellent optical and dielectric properties [30–33]. Further optimization of this growth process requires an understanding of the reaction mechanism. To date, this reaction mechanism remains unclear.

In this work, we intend to elucidate and control this reaction mechanism. To this end, we first investigate the role of the BN precursor by comparing the hBN crystals obtained from PBN and from commercial hBN. From these preliminary results and based on the existing literature, a new theoretical framework for crystal growth is proposed. This hypothetical mechanism is then thoroughly challenged by investigating the influence of several key parameters (composition, temperature and time) and by monitoring the morphology and the (micro-)structure of the grown hBN crystals.

## II. Experimental section

### 1. Synthesis protocol

The precursor preparation following the PDC route has been described elsewhere [34,35]. It includes the borazine synthesis and its polycondensation into PBN at 650°C with 36 wt% $Li_3N$. At this temperature, $Li_3N$ reacts with PBN to form $Li_3BN_2$ [36,37]. Taking into account the PBN [25] and $Li_3N$ [36] weight losses, the composition is $Li_3BN_2$ + 20 wt% PBN. Samples made outside the PDC route (with hBN instead of PBN) are prepared from commercial hBN (Alfa Aesar, ≥ 99.5% purity) and $Li_3BN_2$.

Li$_3$BN$_2$ is synthesized according to the process described by Yamane *et al.* [37] involving the reaction: Li$_3$N + BN → Li$_3$BN$_2$. The Li$_3$N compound (Sigma Aldrich, ≥ 99.5% purity) is mixed with commercial hBN powder in a 1.1:1 molar ratio to compensate for Li$_3$N sublimation. The synthesis is performed at 700°C under an inert atmosphere (Ar flow) for 1h in a molybdenum crucible.

Molybdenum crucibles (inner volume of 433 mm$^3$) are filled with the powder mixture, sealed, and placed in the center of the PCF which is consequently purged several times with argon to remove air contamination. The PCF heating rate is set at 100°C/min and pressure is increased at 3 MPa/min. The target temperature and pressure are thus reached simultaneously. The sample is cooled naturally after the dwell step. The natural cooling of the PCF is rather fast (approximately -150°C/min over the first 400°C).

All of the sample preparation and crucible sealing are performed in a glovebox under argon. Both the Mo crucible and the commercial hBN powder are previously treated under vacuum and then under Ar + 5% H$_2$ at 1200°C for 2h to remove surface contamination.

2. Characterization techniques

After PCF treatment, the sealed crucibles are recovered and opened with a saw. Structural and microstructural analyses of the raw samples are performed by means of powder X-ray diffraction (XRD) performed on a Bruker D8 Advance diffractometer with a Cu K$_\alpha$ source (λ = 1.54060 Å, K$_\beta$ filtered), X-ray tomography (XRT) recorded on an EasyTom Nano (Rx Solutions) operating at 60 kV and 130 μA (LaB$_6$ source), and scanning electron microscopy (SEM) performed with a FEG-SUPRA Zeiss 55VP operating at 1 kV (working distance between 3 mm and 5 mm).

The Li$_3$BN$_2$ present in the PCF-treated samples is dissolved in ultrasonicated hot purified water and the individual crystals are retrieved by a filtration step. Optical microscopy (OM) is performed on an Olympus BX60. The crystals' surface area distributions are obtained from a population of crystals dispersed with ethanol on a flat and transparent surface (Petri dish). They are measured from large area (50 mm*50 mm) OM images. These specific images are acquired on the numerical optical microscope Hirox RH2000 equipped with a binocular module (MXB-2016Z x160). The lighting is crucial for the subsequent image analysis, and has been optimized with a semiannular lighting and a light diffuser. The pixel resolution is 4 μm*4 μm.

The optical properties of an individual crystal are investigated by Raman and Cathodoluminescence (CL) spectroscopy. Raman spectra are recorded on a Labram HR800 spectrometer (HORIBA Jobin-Yvon) with a green laser (532 nm). CL is performed on a homemade setup at Institut Lumière Matière (Lyon, France) described elsewhere [38,39] with a 15 kV acceleration voltage and a source current of 32 μA. The electron beam is not corrected so that its diameter on the surface is roughly 1 mm so that the current density on the sample surface is around 10 μA/mm² (*i.e* $10^{-5}$ pA/nm²). These acquisition conditions are significantly different than in a conventional SEM-integrated CL system where the excitation is very localized, leading to a current density 7 orders of magnitude higher (750 pA/nm² [31]).

The hBN crystal structure is investigated at the atomic scale from transmission electron microscopy (TEM) performed on a JEOL 2100F operating at 200 kV used in high resolution TEM (HRTEM) and selected area electron Diffraction (SAED) modes. The TEM samples are prepared by exfoliation of BNNS from hBN crystals. The exfoliation is achieved by a conventional scotch-tape method [18] using PDMS tape. The exfoliated BNNS is first deposited on a SiO₂/Si wafer and then transferred onto the TEM grid (C-FLAT Protochips) by contact *via* an ethanol droplet. Prior to observation, the TEM grid is treated at 150°C in Ar for 1h for removing polymer and environmental contamination.

### III. Results and discussion

The process we investigate in this article is based on a PDC process coupled to a PCF treatment. However, to elucidate the reaction mechanism, adaptation of this process using commercial hBN powder has been studied.

#### 1. Reactions involved in the hBN crystal growth

Following the PDC route, the molecular structure of the BN precursor is crucial, and the influence of this molecular structure is first investigated. The PDC route applied to PBN supposedly involves a continuous polycondensation of borazine (Figure 1) that is thermally activated [40].

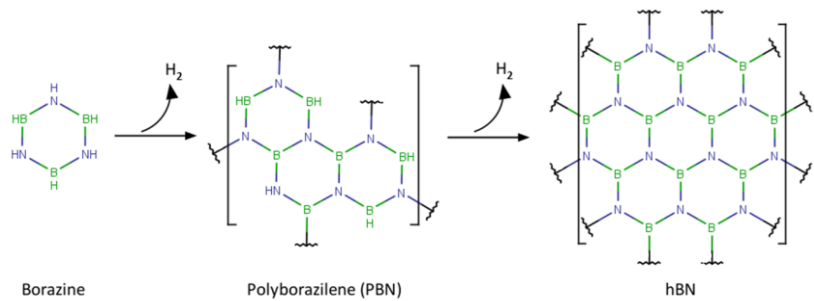

*Figure 1: Hypothetical reaction mechanism of the PDC route applied to hBN synthesized from borazine.*

PBN dehydrogenation would create active sites that should enable further polycondensation [24,27,40–42]. In the framework of the PDC route, the chemical structure of PBN is crucial.

The synthesis process involves an intermediate compound $Li_3N$, known to favor the synthesis of hBN crystallites from PBN [25,27,28]. During the pretreatment, $Li_3N$ and PBN form $Li_3BN_2$ [36,37] which is a compound used as a solvent of hBN in the HPHT process [43–46]. In case of PBN dissolution in $Li_3BN_2$, its chemical structure would be broken. This would imply that the crystallization reaction would occur outside the PDC route.

To determine whether the reaction mechanism occurs inside or outside the PDC route, the process is performed with a commercial hBN powder instead of PBN (1800°C with a 2-hours residence time). The sample composition outside the PDC route is $Li_3BN_2$ + 20 wt% hBN.

XRD patterns (Figure 2a) reveal that both samples are almost exclusively composed of hBN and $Li_3BN_2$. A peak detected at 33.5° is systematically reported in the literature alongside $Li_3BN_2$ [36,37]. This peak is attributed to $Li_2O$ and is supposedly due to the high reactivity of $Li_3BN_2$ with air and moisture [36,37]. Various additional peaks are often detected with a low intensity that can be attributed to various oxides ($Li_xB_yO_z$) (particularly in the hBN sample (Figure 2a) or in the 1800C-15%-2h sample (Figure 5a)).

The relative intensities of the hBN (002) and the $Li_3BN_2$ peaks differ between both samples and would suggest that the proportion of the hBN phase is higher in the PBN sample. This difference is not confirmed by XRT (Figure 2f and g). As the volume probed by XRT is considerably larger than the volume probed by XRD, we assume that this difference reflects sample inhomogeneity.

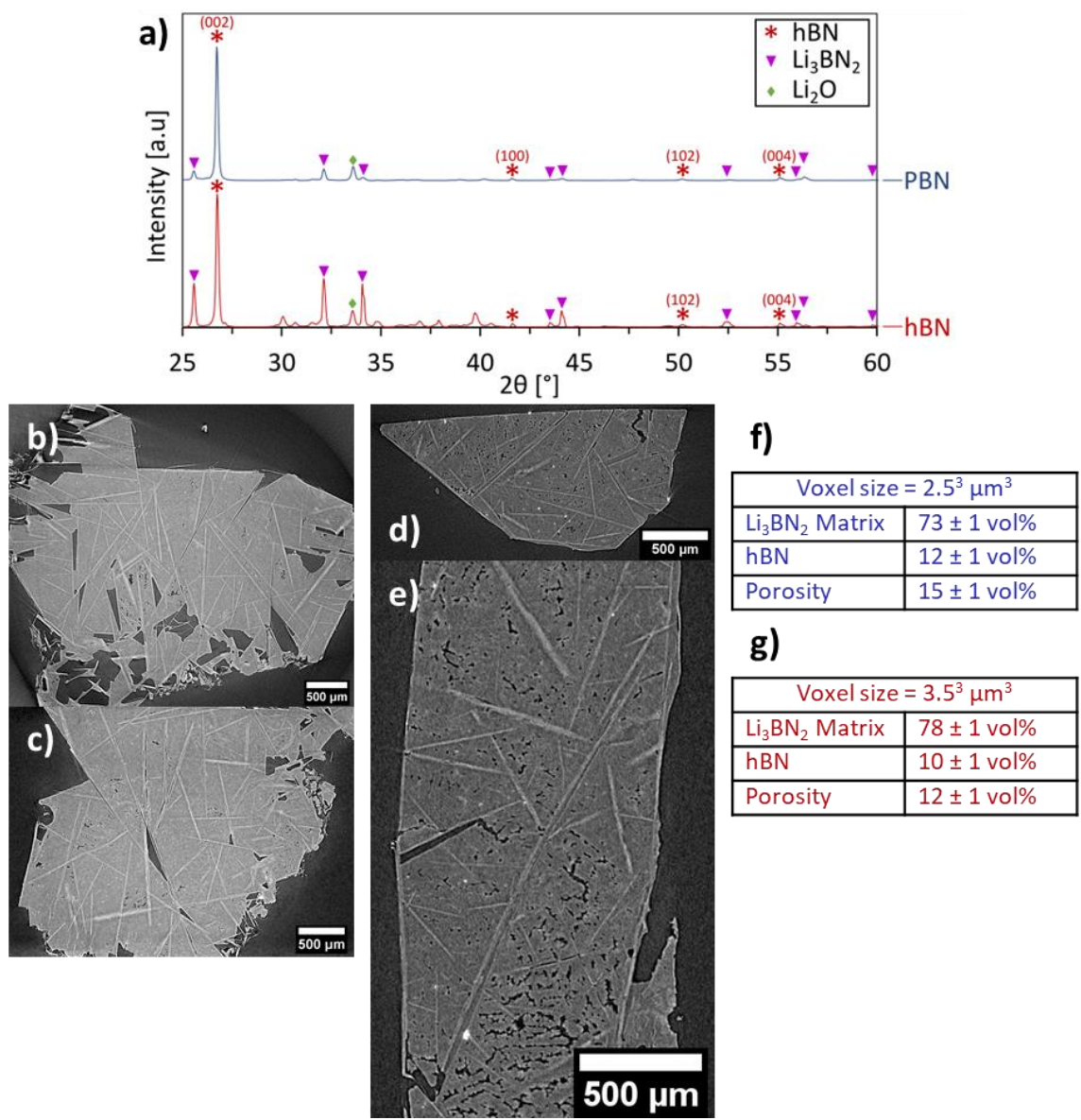

*Figure 2: a) XRD patterns of BN samples prepared with homemade PBN (blue) and commercial hBN (red). b-c) Representative horizontal (b) and vertical (c) slices extracted from the XRT volume (available in Supporting Information) of the sample made from homemade PBN. d-e) Representative horizontal (d) and vertical (e) slices extracted from the XRT volume of the sample made from commercial hBN. White color corresponds to the hBN phase, gray to the $Li_3BN_2$ phase and black to the void. f-g) Tables showing the volume of each phase derived from image analysis for the samples made from (f) homemade PBN and (g) commercial hBN. The uncertainty is derived from subvolume variations. Both samples have been PCF-treated at 1800°C with a 2-hours residence time.*

XRT images (Figure 2b-e and videos 1 and 2 in the Supporting Information) show white lines intertwined in a darker environment. Based on the lamellar morphology of hBN crystals and from the XRD measurements, we attribute the white lines to hBN crystals and darker gray to $Li_3BN_2$. Empty space appears black in XRT. The hBN crystals therefore seem embedded in a large volume of $Li_3BN_2$. Crystal entanglement would indicate that the crystal density is too high (and thus suboptimal).

The as-synthesized samples are generally composed of one monolithic ingot, as in the case of the PBN sample (Figure 2b-c). However, the samples often break into smaller pieces (Figure 2d-e). These pieces are faceted and show shiny surfaces. The SEM view of one of these blocks (Figure SI.1) shows that its surface is indeed covered by hBN, in agreement with XRT. The hBN-covered surfaces seem to result from a mechanical cleavage that preferentially occurs within the crystals, between the hBN layers. This cleavage may be evidence of the weak van der Waals bonding between the hBN layers. The interface between the $Li_3BN_2$ matrix and hBN would thus be significantly stronger. The mechanical cleavage often occurs naturally and could therefore be induced by strain in the sample. This strain could probably be due to contraction during cooling.

The microstructure deduced from XRT indicates that the hBN crystals are present randomly throughout the sample and are not attached to the crucible walls, we therefore assume that there is no preferential growth direction and that the walls do not act as nucleation sites. If heterogeneous nucleation occurs, then it may rely on randomly dispersed nuclei.

Overall, the microstructures of the samples made from PBN and from commercial hBN are very similar, as are the volumes of the different phases inside each sample. This similarity suggests that both samples underwent the same reaction events : the reaction seems to occur outside the PDC route.

In the following, we thus focus exclusively on samples made from commercial hBN as starting material and we assume that the molecular structure of PBN may not be crucial. A new theoretical framework is therefore needed to describe the hBN synthesis under PCF conditions.

2. Toward a solution growth mechanism

To establish this new framework, existing literature can give useful insights. In HPHT conditions, the cBN synthesis occurs by segregation from a solution [12,43–46] and has in particular been demonstrated using $Li_3BN_2$ as solvent [12]. The pressure can selectively determine the formation of the cubic or hexagonal phases, with a threshold around 4.5 GPa for hBN to cBN formation, with BaBN [13] or $Li_3BN_2$ [44] as a solvent. By analogy with these works we can thus infer that hBN crystals may also grow by segregation from a $Li_3BN_2$-BN solution, with a pressure 2 orders of magnitude lower. The structure revealed by XRT may indeed be consistent with a solution growth in a large volume of $Li_3BN_2$.

A supplementary argument in favor of the solution growth mechanism can be found in the hBN crystal morphology (Figure 3) that can be indicative of mechanisms occurring during the growth. Some dendritic-like structures can be found in the crystals synthesized at 1800°C (Figure 3a), but the most spectacular appear when the synthesis process is performed at lower temperature (*cf.* Table 1 for extensive experimental details).

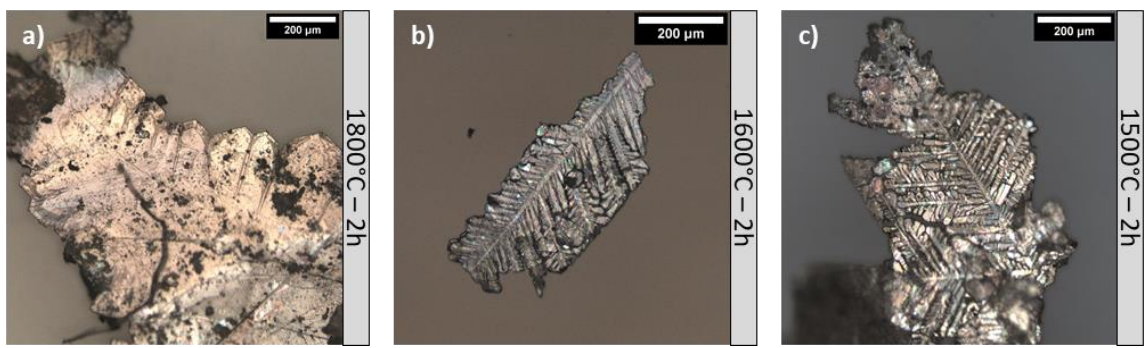

*Figure 3: Optical micrographs of hBN crystals, selected on the basis of the presence of dendrites among a large population. The syntheses are performed at a) 1800°C , b) 1600°C and c) 1500°C and 2h residence time with otherwise identical conditions (cf. Table 1 for experimental details). The 1800°C treatment produces large (~100 µm) dendritic-like structures on the crystal edge while thinner structure is observed after syntheses at 1600°C and 1500°C, with similar secondary interdendritic spacing (~14 µm).*

Some remarkable specimens that were produced at 1600°C (respectively 1500°C) (Figure 3b) (respectively Figure 3c) show dendrites that make up most of the crystal. The primary dendrites are several hundred micrometers long. Secondary (resp. tertiary) dendrites emanate from the primary (respectively secondary) dendrites with a consistent 120° angle (reflecting the hexagonal structure) and a secondary interdendritic spacing of roughly 14 microns.

The dendritic growth has been widely documented [47] and is representative of a solidification mechanism involving a disturbed liquid-solid (or gas-solid) growth interface. The presence of such structures seems to be additional evidence of the hBN crystal growth by segregation from a liquid solution. Moreover, this mechanism seems to be the same for the process temperatures ranging from 1500°C to 1800°C. Generally, dendritic growth is indicative of a fast crystal growth.

We therefore hypothesize that hBN grows by segregation from a $Li_3BN_2$-BN solution. In the following, we question this hypothesis by varying the potentially crucial parameters. The hypothetical solution growth by segregation can traditionally be described in three (or four) successive steps:

1) *(Optional)* Solvent melting;
2) Dissolution;
3) Nucleation;
4) Growth.

The $Li_3BN_2$ melting should occur during the heating phase at 870°C [48]. This melting temperature does not depend on pressure up to 1 GPa [49]. The structure of the $Li_3BN_2$ melt is unclear yet supposedly composed of BN molecules surrounded by Li ions [49]. Upon further heating and during the dwell step, hBN should dissolve into liquid $Li_3BN_2$. Nucleation can be either homo- or hetero- geneous. The heterogeneous nucleation occurs in the presence of heterogeneous nuclei, presumably undissolved hBN or crucible walls. In the absence of a favorable nucleation site, homogeneous nucleation would take place and would be mainly influenced by the driving force for crystallization. At this point, the crystallization driving force is not demonstrated.

The Li$_3$BN$_2$-BN phase equilibrium has been investigated at high pressure (over 4 GPa) [45,50–52] but, to our knowledge, no data is available at lower pressure. The more precise phase diagram was established from thermal analyses and *in situ* synchrotron energy dispersive XRD at 5.3 GPa by Solozhenko *et al.* [45]. It is reproduced and adapted in Figure 4. It is important to emphasize once again that the data from Solozhenko *et al.* were acquired with a pressure 2 orders of magnitude higher than the PCF one. Nonetheless, based on previous works [13,44], the pressure seems to affect the BN crystalline structure without drastically modifying the phase equilibria. In particular, Solozhenko *et al.*'s diagram displays a Li$_3$BN$_2$ peritectic decomposition (P point in Figure 4) at 1347°C. The authors insist on the fact that the peritectic decomposition of Li$_3$BN$_2$ is accompanied by hBN formation in place of cBN [45] and that hBN is metastable in the whole temperature range, cBN being only observed when segregated from the liquid phase, upon cooling. The extrapolation of their data up to 1800°C (red lines in Figure 4) results in an equilibrium liquid composition of approximately 18 wt% hBN.

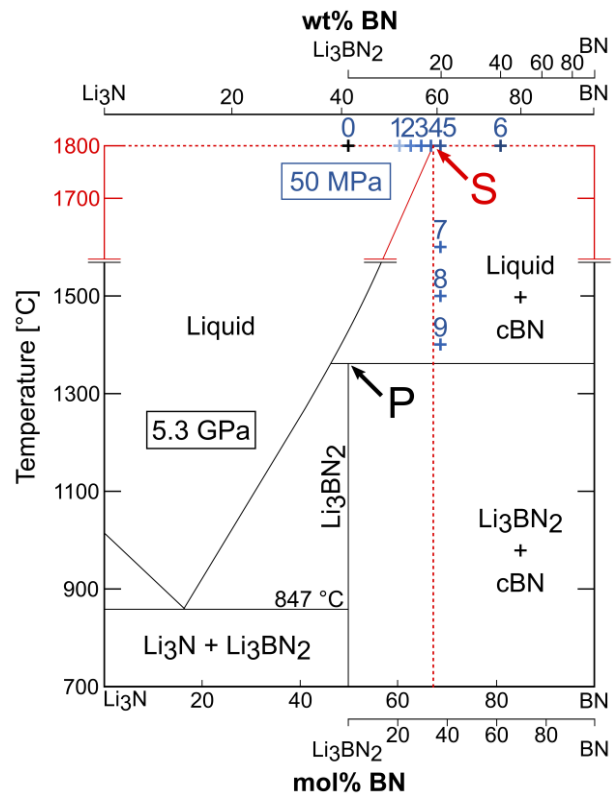

*Figure 4: Li$_3$N-BN phase diagram at HPHT (5.3 GPa) [45] with an extrapolated portion from 1577°C to 1800°C (in red). Experimental data points of the present work acquired under PCF conditions (50 MPa) are represented by blue crosses. The extrapolated BN solubility limit at 1800°C (S point) equals 18 wt% in the Li$_3$BN$_2$-BN system. Reproduced and adapted from ref. 45 (copyright (1997)) with permission from Elsevier.*

By analogy with the HPHT phase equilibria, we can infer that BN solubility in $Li_3BN_2$ increases with the temperature, even under PCF conditions. Undercooling is therefore supposed to be the driving force of crystallization. As we expect no significant temperature gradient in the PCF furnace, growth takes place supposedly during cooling.

At the dwell temperature, equilibrium is supposedly reached when the dissolution limit is reached or when all of the BN has been consumed. The HT dwell time should ensure that this equilibrium can be reached. Overall, the BN dissolution reaction seems to be crucial for the solution growth process. In the following, we thus investigate this reaction mechanism by exploring various process parameters. We explore several initial hBN concentrations and various HT dwell times and temperatures. The parameters explored are summarized in Table 1, and are aggregated to the HPHT phase diagram (blue crosses in Figure 4) for further discussion.

*Table 1: Summary of the samples and synthesis parameters investigated in this paper. The pressure is 50 MPa for all samples (except for PDC 180 MPa [29,30]).*

| Nomenclature | Initial BN concentration versus $Li_3BN_2$ | Dwell temperature | Dwell time at high temperature | Presence of hBN recrystallization | Cross number Figure 4 |
|---|---|---|---|---|---|
| PDC | 20 wt% [29,30] | 1800°C | 2h | Yes | |
| 1800C-0%-2h | 0 wt% | 1800°C | 2h | No | 0 |
| 1800C-10%-2h | 10 wt% | | | | 1 |
| 1800C-12.5%-2h | 12.5 wt% | | | | 2 |
| 1800C-15%-2h | 15 wt% | | | Yes | 3 |
| 1800C-17.5%-2h | 17.5 wt% | | | | 4 |
| 1800C-20%-2h | 20 wt% | | | | 5 |
| 1800C-40%-2h | 40 wt% | | | No | 6 |
| 1600C-20%-2h | 20 wt% | 1600°C | 2h | Yes | 7 |
| 1500C-20%-8h | | 1500°C | 8h | | |
| 1500C-20%-4h | | | 4h | | |
| 1500C-20%-2h | | | 2h | | 8 |
| 1500C-20%-10min | | | 10 min | | |
| 1400C-20%-2h | | 1400°C | 2h | | 9 |

3. BN dissolution-recrystallization reaction in $Li_3BN_2$ at 1800°C

XRD patterns of PCF samples prepared with initial hBN concentration ranging from 0 wt% to 40 wt% (Figure 5) reveal a significant effect of the initial hBN concentration. As expected, the sample made with no initial hBN (1800C-0%-2h) shows only peaks attributed to $Li_3BN_2$ (and $Li_2O$ [36,37]).

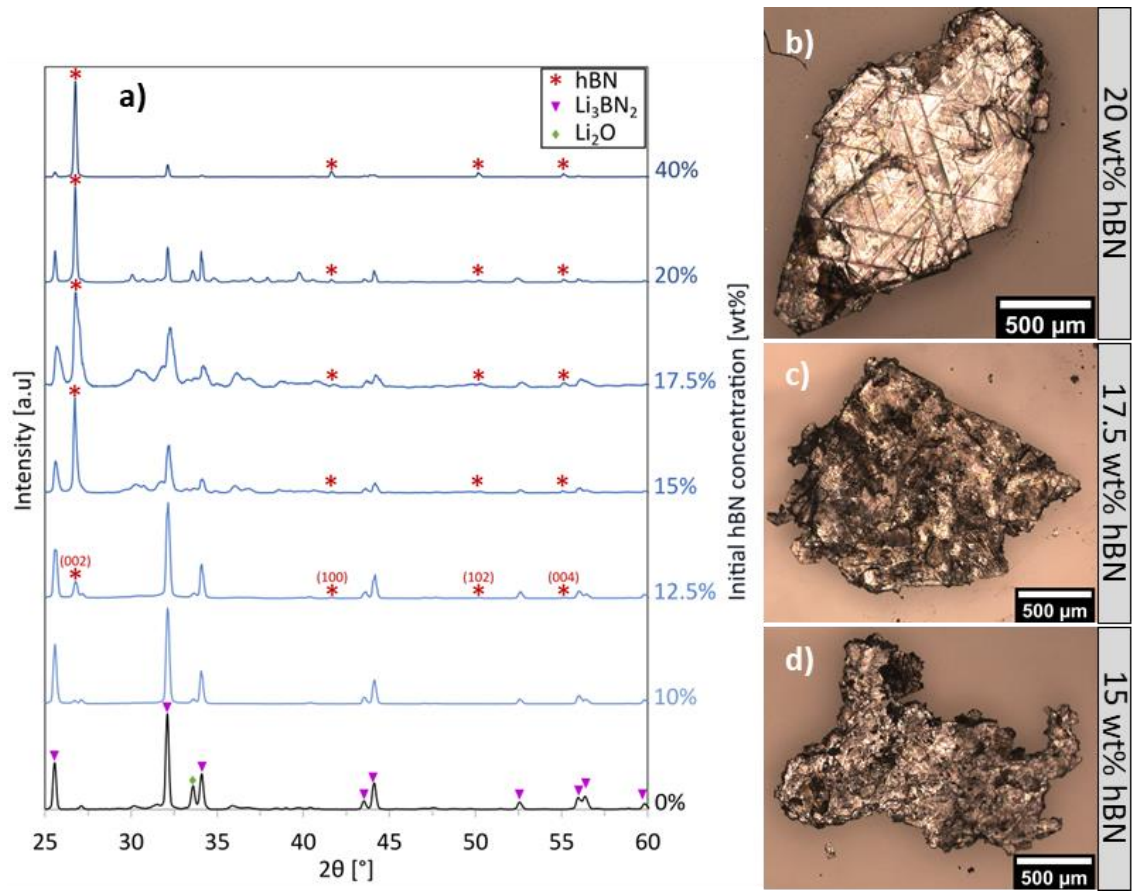

*Figure 5: a) Evolution of the XRD patterns for samples made by PCF treatment (1800°C/2h) as a function of the initial hBN concentration (from bottom to top : 0 wt% to 40 wt%). b)-d) Optical micrographs of hBN crystals obtained for (b) 1800C-20%-2h, (c) 1800C-17.5%-2h and (d) 1800C-15%-2h samples.*

The sample with 10% hBN (1800C-10%-2h) shows almost only peaks attributed to $Li_3BN_2$, with a very low intensity peak detected at 26.7° representative of hBN (002). The PCF-treatment made the hBN phase XRD contribution disappear. When increasing the initial hBN concentration to 12.5% (1800C-12.5%-2h) the XRD spectrum shows a slightly more intense hBN (002) peak at 26.7° which is still very low compared to the expected intensity (see Figure SI.2 for reference patterns). Furthermore, no sample residue is found after dissolution in hot water, which tends to indicate that hBN is almost totally dissolved and does not crystallize subsequently.

The initial hBN concentrations of 15 wt%, 17.5 wt% and 20 wt% produce remarkably different samples. The hBN (002) peak at 26.7° is detected at a much higher intensity compared to $Li_3BN_2$ peaks and millimeter-sized hBN crystals are obtained (representative examples are shown in Figure 5b)-d)). The presence of such crystals may indicate that a dissolution reaction occurs, followed by hBN crystallization.

The PCF-treated 40 wt% hBN sample exhibits an extremely intense hBN (002) XRD peak compared to the peaks associated with $Li_3BN_2$. The sample is powdery and composed of hBN crystallites with morphologies similar to those of the initial commercial hBN powder (Figure SI.3) and some solidified Li$_3$BN$_2$ agglomerates around. This sample therefore appears to be mainly composed of unreacted hBN powder.

We thus conclude that the BN dissolution limit may lie between 12.5 wt% and 40 wt% hBN. The observations on the influence of initial hBN concentration are summarized in Table 2.

*Table 2: Summary of the effect of initial concentration on the final composition and morphology of products from the hBN-Li$_3$BN$_2$ system after PCF treatment at 1800°C/50 MPa/2h.*

| Nomenclature | Initial hBN concentration versus Li$_3$BN$_2$ | hBN XRD peak at 26.7° | Morphology | Interpretation |
|---|---|---|---|---|
| 1800C-0%-2h | 0 wt% | Undetected | ∅ | No spontaneous hBN formation from Li$_3$BN$_2$ |
| 1800C-10%-2h | 10 wt% | Extremely low | ∅ | Complete dissolution & no crystallization |
| 1800C-12.5%-2h | 12.5 wt% | Very Low | ∅ | Complete dissolution & no crystallization |
| 1800C-15%-2h | 15 wt% | Major | Large crystals, very defective | Partial dissolution & recrystallization |
| 1800C-17.5%-2h | 17.5 wt% | Major | Large crystals, less defective | Partial dissolution & recrystallization |
| 1800C-20%-2h | 20 wt% | Major | Large crystals, well organized | Partial dissolution & recrystallization |
| 1800C-40%-2h | 40 wt% | Dominating | Micrometer-sized crystallites | Mainly unreacted commercial powder |

The near-absence of hBN formation in the samples made with 10% (1800C-10%-2h) and 12.5% (1800C-12.5%-2h) initial hBN is particularly remarkable as no stable solid solution in the Li$_3$BN$_2$-BN system is expected [43–46,49,51,53]. The BN structure and location are unclear. We assume that BN segregates out of Li$_3$BN$_2$ at a temperature too low to organize and that it remains amorphous. This phenomenon that needs to be clarified with further studies may be promoted by the fast cooling occurring in the PCF treatment.

We assume that the sharp difference between the 1800C-12.5%-2h and the 1800C-15%-2h samples comes from the nucleation step: nucleation seems to be energetically unfavorable in the 1800C-12.5%-2h sample and favorable in the 1800C-15%-2h sample. However, it is not yet possible to conclude whether nucleation happens homogeneously or heterogeneously. It is, therefore, necessary to take both scenarios into account. Homogeneous nucleation in the solution could occur if all of the BN is dissolved (dissolution limit not reached), and heterogeneous nucleation could occur due to the presence of undissolved heterogeneous hBN nuclei (dissolution limit reached).

Interestingly, the morphology of the crystals is significantly affected by the hBN concentration (Figure 5b)-d)). The crystals produced by the sample 1800C-20%-2h (Figure 5b) show flat surfaces crossed by *striae* [14,15,21] and their edges are rather sharp and geometric. This sharp aspect is also observed in the crystals produced by the 17.5 wt% hBN sample (Figure 5c) but their surface is more defective. Strikingly, the millimeter-sized crystals produced by the 15 wt% hBN sample (Figure 5d) appear very defective with rough surfaces and edges.

Based on the previous results, the initial 20 %wt hBN concentration is selected for further investigation. In the next section, we explore the influence of the time and temperature parameters of the HT dwell.

4. Influence of HT dwell parameters on $Li_3BN_2$-BN solution

By analogy with HPHT studies, the BN dissolution limit in $Li_3BN_2$ supposedly increases with temperature [45,51]. The dwell temperature may thus influence the nucleation step and therefore the size of the grown crystals.

To investigate this effect, we produced samples equivalent to 1800C-20%-2h with dwell temperatures ranging from 1400°C to 1800°C. It appears that the dwell temperature strongly influences the size of the grown crystals (Figure 6a-d). More precisely, the maximum crystal size seems to increase with the dwell temperature.

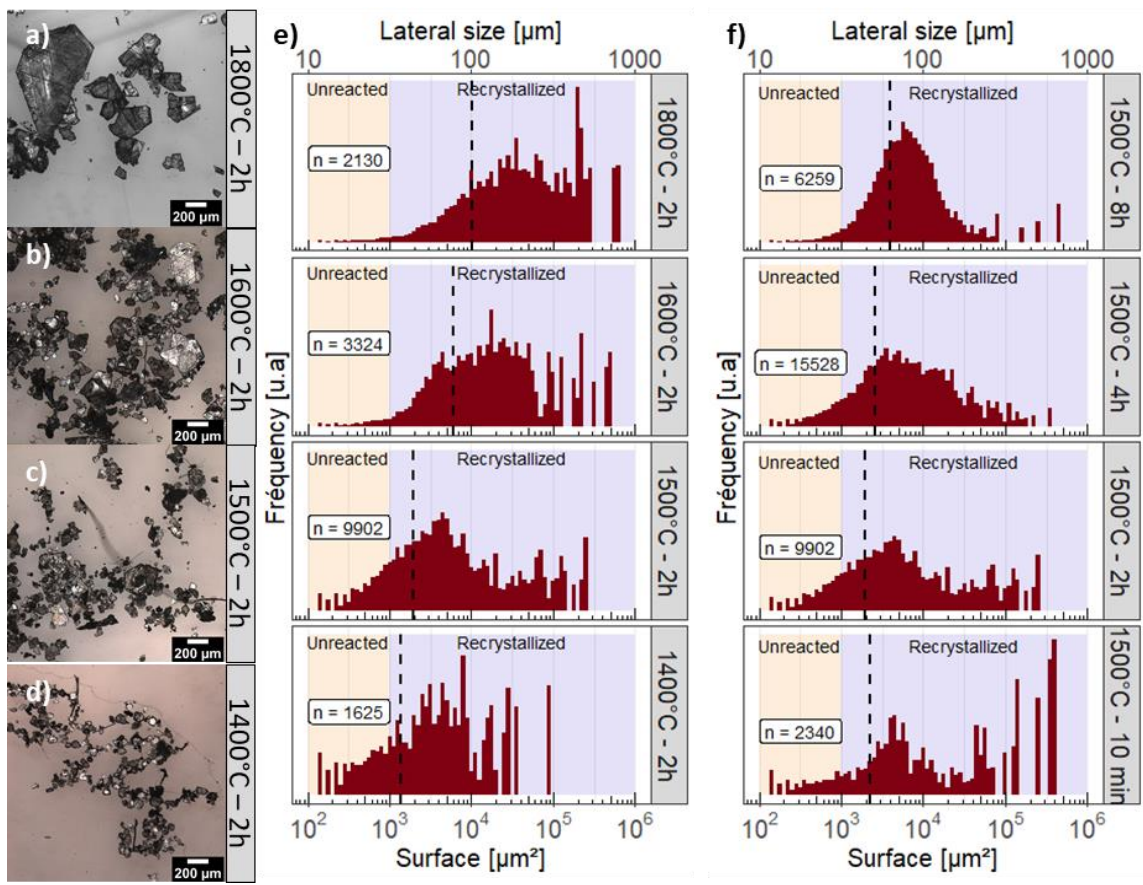

*Figure 6: a-d) Optical microscopy images of crystals synthesized from Li$_3$BN$_2$-20 wt% hBN system at dwell temperatures ranging from 1400°C to 1800°C. e) corresponding weighted surface area distributions; f) surface area distributions of crystals synthesized at 1500°C with a HT dwell duration ranging from 10 min to 8 hours. The mean surface area of each distribution is represented by a dashed line, and the number "n" of crystals per population is inserted.*

Crystal size distributions (Figure 6e) reveal that the dwell temperature influences the entire population of crystals. The temperature increase (from 1400°C to 1800°C) tends to reduce the population with a surface area smaller than $10^3$ µm² (Figure 6e), while increasing the proportion of crystals displaying a surface area larger than $10^5$ µm². Both the mean size and the maximum surface area are thus shifted toward higher values. Moreover, the continuous population shift with temperature appears monotonic. This behavior may indicate that the nucleation mechanism is the same for all of these process temperatures.

The presence of recrystallized hBN in the 1400C-20%-2h sample shows that hBN crystal growth in the Li$_3$BN$_2$-BN system can be achieved at a temperature as low as 1400°C. This particular sample could be decisive to conclude on the nucleation mode.

If nucleation would occur homogeneously, it would mean that all the hBN would be dissolved, in particular at 1400°C. This scenario seems also to contradict studies of the Li$_3$BN$_2$-BN system in HPHT conditions [45,51] which claim that the BN dissolution limit would drastically decrease with temperature from 1800°C to 1400°C (and could even be approaching 0 % at 1400°C).

If nucleation would occur heterogeneously, the lower BN dissolution limit at 1400°C compared to 1800°C would cause a higher density of undissolved hBN. Higher nucleation density would therefore cause an enhanced competition between growing crystals and, consequently, smaller sizes. This scenario seems more consistent with studies of the Li$_3$BN$_2$-BN system in HPHT conditions [45,51].

We therefore rule out the homogeneous nucleation hypothesis and we assume that nucleation is heterogeneous. In the following, as we have ruled out that the crucible walls do not act as nuclei (see section III.1), we consider the heterogeneous nuclei to be only undissolved hBN crystallites.

As a consequence, we can infer that the BN solubility limit in Li$_3$BN$_2$ at 1800°C lies between 12.5 wt% (crystallization would not take place due to a lack of nuclei) and 15 wt% (where significant crystal growth would be enabled by the presence of nuclei).

In the framework of heterogeneous nucleation, the state of the reaction medium at the onset of the growth is crucial. Yet, at the beginning of the PCF treatment, we expect the reaction medium to be highly inhomogeneous due to the preparation process. This inhomogeneity includes the nonuniform powder size distribution (both for hBN and Li$_3$BN$_2$) and their random spatial distribution. The mass transfer mechanisms that could make it more homogeneous (diffusion, convection, particle dispersion, …) are supposedly thermally activated. We investigate these mechanisms by monitoring the size distribution according to the 1500°C dwell duration (Figure 6f). This temperature was selected on the basis of its wide distribution (Figure 6e).

The specific 10-min dwell duration treatment is designed to be as short as possible, but long enough to ensure that the sample is thermalized. This sample produces the widest crystal size distribution: a significant crystal population displaying a large surface area (>$10^5$ µm²) coexisting with unreacted hBN (<$10^3$ µm²). Due to sample preparation, the reaction medium may initially be strongly heterogeneous. For the 1500C-20%-10min sample, the residence time is probably too short for mass transfer mechanisms to happen. Without mass transfer, the medium would remain heterogeneous at the onset of the growth. This heterogeneity seems to lead to a wide crystal size distribution.

The 2-hours dwell treatment produces a thinner distribution, with fewer unreacted crystallites and fewer large crystals. The surface area distribution continuously narrows when the dwell duration increases. Strikingly, the 1500C-20%-8h sample displays a quasi-Gaussian distribution centered on its mean surface area (4 x $10^3$ µm²).

We believe that this behavior would indicate that the reaction medium is slowly homogenized during the dwell step by different hypothetical mechanisms: diffusion in the liquid could homogenize the BN concentration in the solution and the undissolved hBN crystallites could be dispersed. In particular, the quasi-Gaussian distribution of the 1500C-20%-8h sample may indeed originate from a uniform spatial distribution of the nuclei. Additionally, this specific sample seems to be the result of the growth from a quasi-Gaussian nuclei size distribution. To account for this distribution, we assume that the undissolved hBN crystallites could undergo an Oswald ripening mechanism.

5. Summary and overview

Based on the results presented in this work, we propose a schematic scenario for the hBN crystal growth in the $Li_3BN_2$-BN system in PCF conditions (Figure 7). During the heating step $Li_3BN_2$ melts around 870°C and BN dissolution consequently starts, independently of the BN precursor structure (either PBN or hBN).

During the dwell step, BN dissolution continues, and the system is continuously homogenized. This homogenization includes the spatial homogenization of the BN concentration in the liquid and of the undissolved BN particles distribution, but also the homogenization of their size (ripening). These undissolved BN particles should act as nuclei for the subsequent hBN crystal growth during cooling, driven by segregation from a liquid solution (BN solubility decrease).

In this scenario, the dwell step is particularly crucial for the control of crystal size, as it holds homogenization and ripening mechanisms and therefore influences the nucleation step.

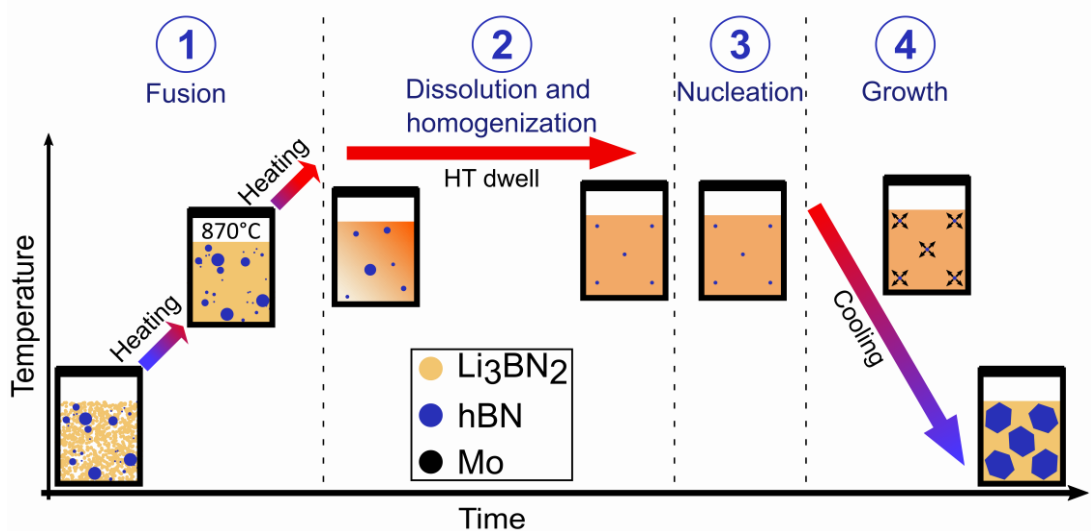

*Figure 7: Summary of the mechanisms involved in the hBN crystal growth in the Li$_3$BN$_2$-BN system. (1) Li$_3$BN$_2$ melts during the heating phase, around 870°C, and hBN starts dissolving subsequently. (2) During the HT dwell, dissolution and homogenization mechanisms occur. (3) Crystallization can occur only in the presence of undissolved hBN crystallites acting as heterogeneous nuclei. (4) Their growth is apparently driven by the decrease in BN solubility during cooling.*

Nucleation determines the density of the growing crystals. Several crystals growing in a close vicinity can compete, and this competition can end up limiting their size. The nucleation step therefore indirectly affects the final crystal size. As an example, crystal entanglement identified by XRT (Figure 2) may indicate that the nucleation density is too high. This result reveals potential for further optimization.

The crystal growth rate is crucial for controlling the defect generation and crystal morphology. Based on our interpretation, the crystal growth is controlled by the cooling rate. This cooling is traditionally very fast and could be the cause of dendritic growth (Figure 3). We investigated this parameter by reducing the cooling rate by 2 orders of magnitude, with no significant influence on crystal size (Figure SI.4). The growth therefore appears extremely fast, and the cooling rate should be drastically reduced to precisely control the growth rate.

The influence of the growth atmosphere has also been investigated by reproducing the 1800C-20%-2h sample with unsealed Mo crucibles under argon and N$_2$. Although the presence of a considerable amount of excess N$_2$ could induce a phase equilibrium shift, no significant change was detected compared to that of Ar (in terms of crystal size and properties). Yet, the molybdenum crucible is badly damaged by N$_2$ at high temperature due to nitridation, and argon is therefore preferred. The sealing seems to have no significant effect on crystal growth, it is performed to isolate the sample from carbon contamination due to the furnace.

All of these mechanisms are additionally influenced by the initial BN concentration with respect to the dissolution limit at dwell temperature. The extrapolation of Solozhenko *et al.'s* data up to 1800°C (red lines in Figure 4) results in a dissolution limit of approximately 18 wt% hBN at 1800°C. In the present work, we estimate this limit between 12.5 wt% and 15 wt% at 1800°C. This amount is lower yet close to the extrapolated value. This relative agreement and the mean crystal size evolution with respect to the dwell temperature seem to demonstrate a similarity between the mechanisms unveiled in this paper and the ones at work under HPHT conditions.

6. Structural and optical characterization of the hBN crystals

In the first parts of this *results and discussion* section, we show that the growth mechanism does not depend on the structure of the BN precursor. In previous papers, we reported that hBN crystals produced from PBN demonstrate excellent structural properties [29,30]. In this part, we aim to investigate the influence of the BN precursor on the hBN properties by characterizing crystals extracted from the 1800C-20%-2h sample.

Crystals extracted from the 1800C-20%-2h sample (Figure 8a) are transparent and show triangular monocrystalline domains up to several hundreds of micrometers. It also presents a flat surface by SEM (Figure 8b).

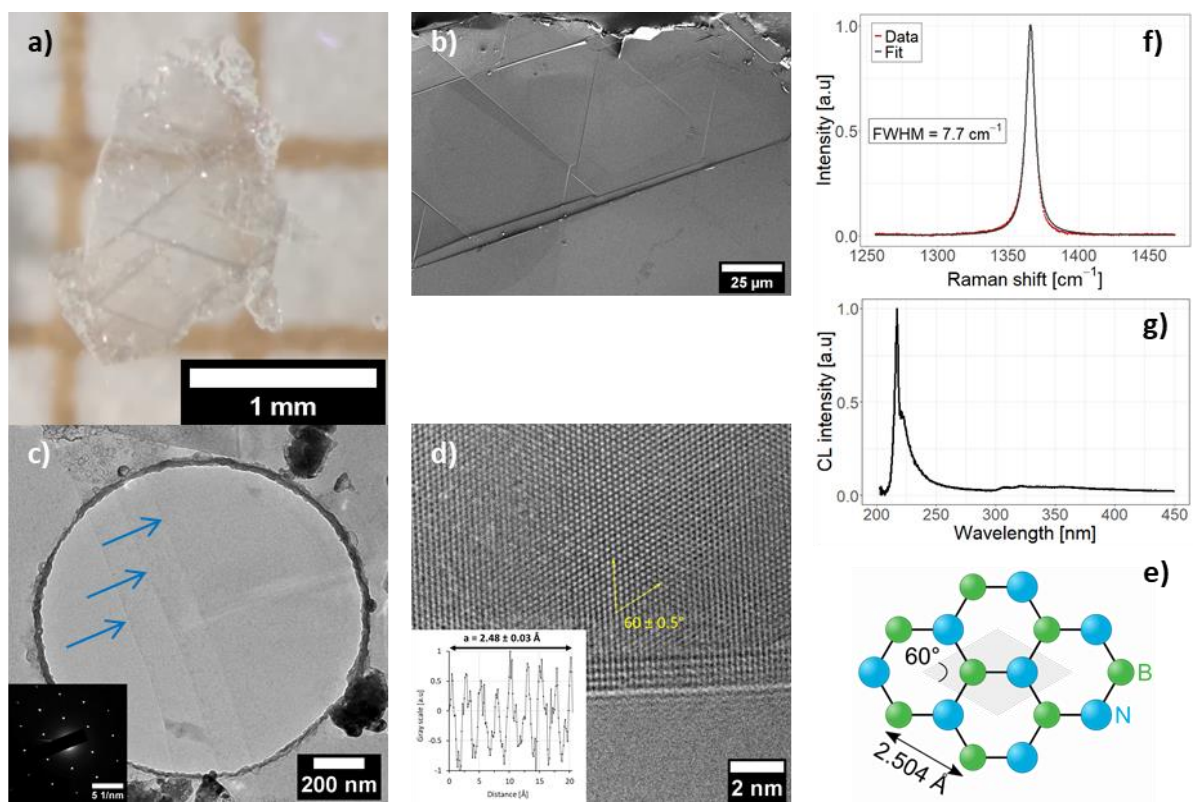

*Figure 8: a) Optical and (b) SEM images of hBN crystal extracted from the 1800C-20%-2h sample and (c) TEM image showing a BNNS exfoliated from this hBN crystal showing different thicknesses (indicated by blue arrows) and a perfect hexagonal structure as probed by SAED (inset). (d) HRTEM image revealing a highly organized hexagonal arrangement displaying a lattice parameter (inset) similar to the (e) expected value. These crystals exhibit state-of-the-art hBN optical properties: (f)*

*Raman E$_{2g}$ peak and its fwhm (inset) and (g) cathodoluminescence spectrum dominated by hBN's intrinsic emission at 215 nm.*

The HRTEM image acquired on a BNNS exfoliated from these crystals (Figure 8c) shows a regular hexagonal structure as confirmed by SAED (inset Figure 8c). The planar lattice parameter derived from HRTEM images (Figure 8d) equals *2.48 ± 0.03 Å* in good agreement with reported values (*2.5038 Å*) [54].

The full width at half maximum (fwhm) of the hBN Raman E$_{2g}$ peak (Figure 8f) is 7.7 cm$^{-1}$, *i.e.*, equivalent to the lowest values measured on natural hBN[55], and would thus indicate a high crystallinity of this material. However, Raman spectroscopy has been shown insufficient for characterizing high quality hBN crystals and luminescence measurement may be preferred [56].

The CL spectrum acquired on one of these crystals (Figure 8g) shows a strong emission at 215 nm. This emission corresponds to the hBN phonon-assisted indirect emission [57]. The presence and the intensity of this intrinsic emission are supplementary signs of structural quality. The structural defects band at 227 nm [56] is also detected with usual intensity [30]. Finally, a contribution structured in quadruplet between 300 nm and 350 nm appears at low intensity. These additional emissions are also detected in HPHT [58], APHT [31], and PDC [30,31] hBN crystals. They are probably a sign of residual doping, presumably carbon [58] yet their origin is still subject to discussions [59]. The PCF furnace is composed of graphite, and carbon doping thus needs to be considered, even if the crucible is sealed. Additionally, the presence of lithium in the crystal structure is probable, but our efforts toward its detection have been unsuccessful.

To conclude, the millimetric crystals from the 1800C-20%-2h sample are actually composed of hBN with monocrystalline domains up to several hundreds of micrometers. They demonstrate high crystallinity and show structural and optical properties very similar to the state-of-the-art freestanding hBN crystals reported in the literature [21].

IV.    Conclusion

The present work demonstrates the hBN crystal growth in the Li$_3$BN$_2$-BN system independently of the BN precursor's structure, *i.e*, outside the PDC route that has been previously proposed. Microstructural analyses are compatible with a scenario of crystal growth by segregation from a Li$_3$BN$_2$-BN solution.

The BN dissolution reaction at 1800°C is investigated by *post-mortem* structural analysis, and three domains of apparent equilibrium are identified. Below 12.5 wt% BN, the dissolution is complete and crystallization does not take place. Between 15 wt% and at least 20 wt% BN, dissolution and recrystallization take place. At 40 wt% BN, the solubility limit is considerably exceeded and most of the hBN remains unreacted. These differences are attributed to the nucleation mechanism.

At a fixed composition, the nucleation mechanism is influenced by the dissolution reaction and so by the PCF process parameters. By monitoring the final crystal size according to the dwell time and temperature parameters, we show that nucleation seems to take place heterogeneously from undissolved hBN crystallites. This conclusion implies that the BN solubility in $Li_3BN_2$ at 1800°C and 50 MPa should lie between 12.5 wt% and 15 wt% hBN. We also identify a homogenization mechanism of the reaction medium during the dwell step with a probable Oswald ripening of the nuclei. These homogenization mechanisms appear dwell temperature and time dependent.

In that, we identify the process parameters crucial to control the final crystal size. A competition effect that may limit the crystal size is also identified, which represents an opportunity for further optimization. Among all the parameters explored, the 1800C-20%-2h sample produces the largest crystals. The crystals extracted from this sample display structural and optical properties similar to those of the state-of-the-art freestanding crystals.

The growth of large hBN crystals from a $Li_3BN_2$-BN system has not been reported before at a moderate pressure. Moreover, significant hBN growth is identified with a dwell temperature as low as 1400°C. These relatively soft conditions could be reproduced in a dedicated flux growth setup. The high solubility of BN in the molten $Li_3BN_2$ bath makes it one of the best solvents for hBN solution growth. Overall, this paper thus demonstrates that the $Li_3BN_2$-BN system is promising for the flux growth of bulk hBN and may pave the way for continuous seeded hBN growth.

### Acknowledgment

The authors thank the CLYM and the CTµ, for providing access to the SEM and TEM facilities, the CECOMO for access to Raman spectroscopy, the ILM, and specifically C. Dujarding for granting access to the cathodoluminescence setup. The authors also thank J. Adrien (MATEIS) who performed the XRT measurements.

The authors also thank the UCBL mechanical workshop and its members for their technical support and manufacturing of the synthesis crucibles.

This work has been partially financially supported by the European Union Horizon 2020 Program under the Graphene Flagship (Graphene Core 3, grant number 881603) and by the French National Research Agency (project ELuSeM n° ANR-21-CE24-0025).

### Supporting information description

In the supporting information, we bring more details on the XRT measurements by showing the full reconstructed volume in a video format and a SEM view of the analyzed sample. We present XRD data acquired on $Li_3BN_2$-hBN non-annealed standards to fortify the phase quantifications discussed in the main text. We compare the morphology of the initial hBN powder and the 1800C-40%-2h sample. Finally, crystal growth rate is briefly investigated based on cooling rate modification.

## References

(1) Geim, A. K.; Grigorieva, I. V. Van Der Waals Heterostructures. *Nature* **2013**, *499* (7459), 419–425. https://doi.org/10.1038/nature12385.

(2) Dean, C. R.; Young, A. F.; Meric, I.; Lee, C.; Wang, L.; Sorgenfrei, S.; Watanabe, K.; Taniguchi, T.; Kim, P.; Shepard, K. L.; Hone, J. Boron Nitride Substrates for High-Quality Graphene Electronics. *Nat. Nanotechnol.* **2010**, *5* (10), 722–726. https://doi.org/10.1038/nnano.2010.172.

(3) Watanabe, K.; Taniguchi, T.; Kanda, H. Direct-Bandgap Properties and Evidence for Ultraviolet Lasing of Hexagonal Boron Nitride Single Crystal. *Nat. Mater.* **2004**, *3* (6), 404–409. https://doi.org/10.1038/nmat1134.

(4) Zhang, Z.; Hu, S.; Chen, J.; Li, B. Hexagonal Boron Nitride: A Promising Substrate for Graphene with High Heat Dissipation. *Int. Heat Transf. Conf.* **2018**, *2018-Augus*, 5181–5189. https://doi.org/10.1615/ihtc16.hte.022180.

(5) Doan, T. C.; Majety, S.; Grenadier, S.; Li, J.; Lin, J. Y.; Jiang, H. X. Fabrication and Characterization of Solid-State Thermal Neutron Detectors Based on Hexagonal Boron Nitride Epilayers. *Nucl. Instruments Methods Phys. Res. Sect. A Accel. Spectrometers, Detect. Assoc. Equip.* **2014**, *748*, 84–90. https://doi.org/10.1016/j.nima.2014.02.031.

(6) Chen, T.-A.; Chuu, C.-P.; Tseng, C.-C.; Wen, C.-K.; Wong, H.-S. P.; Pan, S.; Li, R.; Chao, T.-A.; Chueh, W.-C.; Zhang, Y.; Fu, Q.; Yakobson, B. I.; Chang, W.-H.; Li, L.-J. Wafer-Scale Single-Crystal Hexagonal Boron Nitride Monolayers on Cu (111). *Nature* **2020**, *579* (7798), 219–223. https://doi.org/10.1038/s41586-020-2009-2.

(7) Ma, K. Y.; Zhang, L.; Jin, S.; Wang, Y.; Yoon, S. I.; Hwang, H.; Oh, J.; Jeong, D. S.; Wang, M.; Chatterjee, S.; Kim, G.; Jang, A.; Yang, J.; Ryu, S.; Jeong, H. Y.; Ruoff, R. S.; Chhowalla, M.; Ding, F.; Shin, H. S. Epitaxial Single-Crystal Hexagonal Boron Nitride Multilayers on Ni (111). *Nature* **2022**, *606* (7912), 88–93. https://doi.org/10.1038/s41586-022-04745-7.

(8) Fukamachi, S.; Solís-Fernández, P.; Kawahara, K.; Tanaka, D.; Otake, T.; Lin, Y.-C.; Suenaga, K.; Ago, H. Large-Area Synthesis and Transfer of Multilayer Hexagonal Boron Nitride for Enhanced Graphene Device Arrays. *Nat. Electron.* **2023**, *6* (2), 126–136. https://doi.org/10.1038/s41928-022-00911-x.

(9) Cai, J.; Anderson, E.; Wang, C.; Zhang, X.; Liu, X.; Holtzmann, W.; Zhang, Y.; Fan, F.; Taniguchi, T.; Watanabe, K.; Ran, Y.; Cao, T.; Fu, L.; Xiao, D.; Yao, W.; Xu, X. Signatures of Fractional Quantum Anomalous Hall States in Twisted MoTe2. *Nature* **2023**, *622* (7981), 63–68. https://doi.org/10.1038/s41586-023-06289-w.

(10) Li, Y.; Zhang, F.; Ha, V.-A.; Lin, Y.-C.; Dong, C.; Gao, Q.; Liu, Z.; Liu, X.; Ryu, S. H.; Kim, H.; Jozwiak, C.; Bostwick, A.; Watanabe, K.; Taniguchi, T.; Kousa, B.; Li, X.; Rotenberg, E.; Khalaf, E.; Robinson, J. A.; Giustino, F.; Shih, C.-K. Tuning Commensurability in Twisted van Der Waals Bilayers. *Nature* **2024**, *625* (7995), 494–499. https://doi.org/10.1038/s41586-023-06904-w.

(11) Cording, L.; Liu, J.; Tan, J. Y.; Watanabe, K.; Taniguchi, T.; Avsar, A.; Özyilmaz, B. Highly Anisotropic Spin Transport in Ultrathin Black Phosphorus. *Nat. Mater.* **2024**. https://doi.org/10.1038/s41563-023-01779-8.

(12) Taniguchi, T.; Yamaoka, S. Spontaneous Nucleation of Cubic Boron Nitride Single Crystal by Temperature Gradient Method under High Pressure. *J. Cryst. Growth* **2001**, *222* (3), 549–557. https://doi.org/10.1016/S0022-0248(00)00907-6.

(13) Taniguchi, T.; Watanabe, K. Synthesis of High-Purity Boron Nitride Single Crystals under High Pressure by Using Ba–BN Solvent. *J. Cryst. Growth* **2007**, *303* (2), 525–529. https://doi.org/10.1016/j.jcrysgro.2006.12.061.

(14) Kubota, Y.; Watanabe, K.; Tsuda, O.; Taniguchi, T. Deep Ultraviolet Light-Emitting Hexagonal Boron Nitride Synthesized at Atmospheric Pressure. *Science (80-. ).* **2007**, *317* (5840), 932–934. https://doi.org/10.1126/science.1144216.

(15) Edgar, J. H.; Hoffman, T. B.; Clubine, B.; Currie, M.; Du, X. Z.; Lin, J. Y.; Jiang, H. X. Characterization of Bulk Hexagonal Boron Nitride Single Crystals Grown by the Metal Flux Technique. *J. Cryst. Growth* **2014**, *403*, 110–113. https://doi.org/10.1016/j.jcrysgro.2014.06.006.

(16) Hoffman, T. B.; Zhang, Y.; Edgar, J. H.; Gaskill, D. K. Growth of HBN Using Metallic Boron: Isotopically Enriched h 10 BN and h 11 BN. *MRS Proc.* **2014**, *1635* (3), 35–40. https://doi.org/10.1557/opl.2014.48.

(17) Hoffman, T. B.; Clubine, B.; Zhang, Y.; Snow, K.; Edgar, J. H. Optimization of Ni–Cr Flux Growth for Hexagonal Boron Nitride Single Crystals. *J. Cryst. Growth* **2014**, *393*, 114–118. https://doi.org/10.1016/j.jcrysgro.2013.09.030.

(18) Zhang, X.; Li, Y.; Mu, W.; Bai, W.; Sun, X.; Zhao, M.; Zhang, Z.; Shan, F.; Yang, Z. Advanced Tape-Exfoliated Method for Preparing Large-Area 2D Monolayers: A Review. *2D Mater.* **2021**, *8* (3), 032002. https://doi.org/10.1088/2053-1583/ac016f.

(19) Ouaj, T.; Kramme, L.; Metzelaars, M.; Li, J.; Watanabe, K.; Taniguchi, T.; Edgar, J. H.; Beschoten, B.; Kögerler, P.; Stampfer, C. Chemically Detaching HBN Crystals Grown at Atmospheric Pressure and High Temperature for High-Performance Graphene Devices. *Nanotechnology* **2023**, *34* (47), 475703. https://doi.org/10.1088/1361-6528/acf2a0.

(20) Li, J.; Yuan, C.; Elias, C.; Wang, J.; Zhang, X.; Ye, G.; Huang, C.; Kuball, M.; Eda, G.; Eda, G.; Eda, G.; Redwing, J. M.; He, R.; Cassabois, G.; Gil, B.; Valvin, P.; Pelini, T.; Liu, B.; Edgar, J. H. Hexagonal Boron Nitride Single Crystal Growth from Solution with a Temperature Gradient. *Chem. Mater.* **2020**, *32* (12), 5066–5072. https://doi.org/10.1021/acs.chemmater.0c00830.

(21) Maestre, C.; Toury, B.; Steyer, P.; Garnier, V.; Journet, C. Hexagonal Boron Nitride: A Review on Selfstanding Crystals Synthesis towards 2D Nanosheets. *J. Phys. Mater.* **2021**, *4* (4), 044018. https://doi.org/10.1088/2515-7639/ac2b87.

(22) Colombo, P.; Mera, G.; Riedel, R.; Sorarù, G. D. Polymer-Derived Ceramics: 40 Years of Research and Innovation in Advanced Ceramics. *J. Am. Ceram. Soc.* **2010**, *93* (7), 1805–1837. https://doi.org/10.1111/j.1551-2916.2010.03876.x.

(23) Bernard, S.; Miele, P. Polymer-Derived Boron Nitride: A Review on the Chemistry, Shaping and Ceramic Conversion of Borazine Derivatives. *Materials (Basel).* **2014**, *7* (11), 7436–7459. https://doi.org/10.3390/ma7117436.

(24) Matsoso, B.; Hao, W.; Li, Y.; Vuillet-a-Ciles, V.; Garnier, V.; Steyer, P.; Toury, B.; Marichy, C.; Journet, C. Synthesis of Hexagonal Boron Nitride 2D Layers Using Polymer Derived Ceramics Route and Derivatives. *J. Phys. Mater.* **2020**, *3* (3), 034002. https://doi.org/10.1088/2515-7639/ab854a.

(25) Yuan, S.; Toury, B.; Benayoun, S.; Chiriac, R.; Gombault, F.; Journet, C.; Brioude, A. Low-Temperature Synthesis of Highly Crystallized Hexagonal Boron Nitride Sheets with Li 3 N as Additive Agent. *Eur. J. Inorg. Chem.* **2014**, *2014* (32), 5507–5513. https://doi.org/10.1002/ejic.201402507.

(26) Cornu, D.; Miele, P.; Toury, B.; Bonnetot, B.; Mongeot, H.; Bouix, J. Boron Nitride Matrices and Coatings from Boryl Borazine Molecular Precursors. *J. Mater. Chem.* **1999**, *9* (10), 2605–2610. https://doi.org/10.1039/a904318g.

(27) Yuan, S.; Journet, C.; Linas, S.; Garnier, V.; Steyer, P.; Benayoun, S.; Brioude, A.; Toury, B. How to Increase the H-BN Crystallinity of Microfilms and Self-Standing Nanosheets: A Review of the Different Strategies Using the PDCs Route. *Crystals* **2016**, *6* (5), 55. https://doi.org/10.3390/cryst6050055.

(28) Li, Y.; Garnier, V.; Journet, C.; Barjon, J.; Loiseau, A.; Stenger, I.; Plaud, A.; Toury, B.; Steyer, P. Advanced Synthesis of Highly Crystallized Hexagonal Boron Nitride by Coupling Polymer-Derived Ceramics and Spark Plasma Sintering Processes—Influence of the Crystallization Promoter and Sintering Temperature. *Nanotechnology* **2019**, *30* (3), 035604. https://doi.org/10.1088/1361-6528/aaebb4.

(29) Li, Y.; Garnier, V.; Steyer, P.; Journet, C.; Toury, B. Millimeter-Scale Hexagonal Boron Nitride Single Crystals for Nanosheet Generation. *ACS Appl. Nano Mater.* **2020**, *3* (2), 1508–1515. https://doi.org/10.1021/acsanm.9b02315.

(30) Maestre, C.; Li, Y.; Garnier, V.; Steyer, P.; Roux, S.; Plaud, A.; Loiseau, A.; Barjon, J.; Ren, L.; Robert, C.; Han, B.; Marie, X.; Journet, C.; Toury, B. From the Synthesis of HBN Crystals to Their Use as Nanosheets in van Der Waals Heterostructures. *2D Mater.* **2022**, *9* (3), 035008. https://doi.org/10.1088/2053-1583/ac6c31.

(31) Roux, S.; Arnold, C.; Paleari, F.; Sponza, L.; Janzen, E.; Edgar, J. H.; Toury, B.; Journet, C.; Garnier, V.; Steyer, P.; Taniguchi, T.; Watanabe, K.; Ducastelle, F.; Loiseau, A.; Barjon, J. Radiative Lifetime of Free Excitons in Hexagonal Boron Nitride. *Phys. Rev. B* **2021**, *104* (16), L161203. https://doi.org/10.1103/PhysRevB.104.L161203.

(32) Schmitt, A.; Mele, D.; Rosticher, M.; Taniguchi, T.; Watanabe, K.; Maestre, C.; Journet, C.; Garnier, V.; Fève, G.; Berroir, J. M.; Voisin, C.; Plaçais, B.; Baudin, E. High-Field 1/f Noise in HBN-Encapsulated Graphene Transistors. *Phys. Rev. B* **2023**, *107* (16), L161104. https://doi.org/10.1103/PhysRevB.107.L161104.

(33) Pierret, A.; Mele, D.; Graef, H.; Palomo, J.; Taniguchi, T.; Watanabe, K.; Li, Y.; Toury, B.; Journet, C.; Steyer, P.; Garnier, V.; Loiseau, A.; Berroir, J.-M.; Bocquillon, E.; Feve, G.; Voisin, C.; Baudin, E.; Rosticher, M.; Placais, B. Dielectric Permittivity, Conductivity and Breakdown FIeld of Hexagonal Boron Nitride. *Mater.*

*Res. Express* **2022**. https://doi.org/10.1088/2053-1591/ac4fe1.
(34) Wideman, T.; Sneddon, L. G. Convenient Procedures for the Laboratory Preparation of Borazine. *Inorg. Chem.* **1995**, *34* (4), 1002–1003. https://doi.org/10.1021/ic00108a039.
(35) Yuan, S.; Toury, B.; Journet, C.; Brioude, A. Synthesis of Hexagonal Boron Nitride Graphene-like Few Layers. *Nanoscale* **2014**, *6* (14), 7838–7841. https://doi.org/10.1039/c4nr01017e.
(36) Sahni, K.; Ashuri, M.; Emani, S.; Kaduk, J. A.; Németh, K.; Shaw, L. L. On the Synthesis of Lithium Boron Nitride (Li3BN2). *Ceram. Int.* **2018**, *44* (7), 7734–7740. https://doi.org/10.1016/j.ceramint.2018.01.200.
(37) Yamane, H.; Kikkawa, S.; Koizumi, M. High- and Low-Temperature Phases of Lithium Boron Nitride, Li3BN2: Preparation, Phase Relation, Crystal Structure, and Ionic Conductivity. *J. Solid State Chem.* **1987**, *71* (1), 1–11. https://doi.org/10.1016/0022-4596(87)90135-6.
(38) Hospodková, A.; Oswald, J.; Zíková, M.; Pangrác, J.; Kuldová, K.; Blažek, K.; Ledoux, G.; Dujardin, C.; Nikl, M. On the Correlations between the Excitonic Luminescence Efficiency and the QW Numbers in Multiple InGaN/GaN QW Structure. *J. Appl. Phys.* **2017**, *121* (21), 214505. https://doi.org/10.1063/1.4984908.
(39) Hospodkova, A.; Hubacek, T.; Oswald, J.; Pangrac, J.; Kuldova, K.; Hyvl, M.; Dominec, F.; Ledoux, G.; Dujardin, C. InGaN/GaN Structures: Effect of the Quantum Well Number on the Cathodoluminescent Properties. *Phys. status solidi* **2018**, *255* (5), 1700464. https://doi.org/10.1002/pssb.201700464.
(40) Bernard, S.; Miele, P. Nanostructured and Architectured Boron Nitride from Boron, Nitrogen and Hydrogen-Containing Molecular and Polymeric Precursors. *Mater. Today* **2014**, *17* (9), 443–450. https://doi.org/10.1016/j.mattod.2014.07.006.
(41) Paine, R. T.; Narula, C. K. Synthetic Routes to Boron Nitride. *Chem. Rev.* **1990**, *90* (1), 73–91. https://doi.org/10.1021/cr00099a004.
(42) Termoss, H.; Toury, B.; Brioude, A.; Dazord, J.; Le Brusq, J.; Miele, P. High Purity Boron Nitride Thin Films Prepared by the PDCs Route. *Surf. Coatings Technol.* **2007**, *201* (18), 7822–7828. https://doi.org/10.1016/j.surfcoat.2007.03.016.
(43) Nakano, S.; Ikawa, H.; Fukunaga, O. Synthesis of Cubic Boron Nitride Using Li3BN2, Sr3B2N4 and Ca3B2N4 as Solvent-Catalysts. *Diam. Relat. Mater.* **1994**, *3* (1–2), 75–82. https://doi.org/10.1016/0925-9635(94)90034-5.
(44) Fukunaga, O.; Takeuchi, S. Growth Mechanism of Cubic BN Using Li3BN2 Solvent under High Pressure. *Int. J. Refract. Met. Hard Mater.* **2016**, *55*, 54–57. https://doi.org/10.1016/j.ijrmhm.2015.11.008.
(45) Solozhenko, V. L.; Turkevich, V. Z. High Pressure Phase Equilibria in the Li3N-BN System: In Situ Studies. *Mater. Lett.* **1997**, *32* (2–3), 179–184. https://doi.org/10.1016/S0167-577X(97)00025-6.
(46) Bocquillon, G.; Loriers-Susse, C.; Loriers, J. Synthesis of Cubic Boron Nitride Using Mg and Pure or M'-Doped Li3N, Ca3N2 and Mg3N2 with M'=Al, B, Si, Ti. *J. Mater. Sci.* **1993**, *28* (13), 3547–3556. https://doi.org/10.1007/BF01159836.
(47) D.T.J, H. *Handbook of Crystal Growth. Vol. 1: Fundamentals. a : Thermodynamics and Kinetics*; Elsevier: Amsterdam, 1993. https://doi.org/10.1107/S010876739400512X.
(48) Goubeau, J.; Anselment, W. Uber Ternare Metall-Bornitride. *Zeitschrift fur Anorg. und Allg. Chemie* **1961**, *310* (4–6), 248–260. https://doi.org/10.1002/zaac.19613100410.
(49) DeVries, R. C.; Fleischer, J. F. The System Li3BN2 at High Pressures and Temperatures. *Mater. Res. Bull.* **1969**, *4* (7), 433–441. https://doi.org/10.1016/0025-5408(69)90086-5.
(50) Wentorf, R. H. Synthesis of the Cubic Form of Boron Nitride. *J. Chem. Phys.* **1961**, *34* (3), 809–812. https://doi.org/10.1063/1.1731679.
(51) DeVries, R. C.; Fleischer, J. F. Phase Equilibria Pertinent to the Growth of Cubic Boron Nitride. *J. Cryst. Growth* **1972**, *13–14* (C), 88–92. https://doi.org/10.1016/0022-0248(72)90068-1.
(52) Turkevich, V. Z. Phase Diagrams and Synthesis of Cubic Boron Nitride. *J. Phys. Condens. Matter* **2002**, *14* (44), 10963–10968. https://doi.org/10.1088/0953-8984/14/44/410.
(53) Solozhenko, V. L.; Turkevich, V. Z.; Holzapfel, W. B. Refined Phase Diagram of Boron Nitride. *J. Phys. Chem. B* **1999**, *103* (15), 2903–2905. https://doi.org/10.1021/jp984682c.
(54) Pease, R. S. An X-Ray Study of Boron Nitride. *Acta Crystallogr.* **1952**, *5* (3), 356–361. https://doi.org/10.1107/S0365110X52001064.
(55) Janzen, E.; Schutte, H.; Plo, J.; Rousseau, A.; Michel, T.; Desrat, W.; Valvin, P.; Jacques, V.; Cassabois, G.; Gil, B.; Edgar, J. H. Boron and Nitrogen Isotope Effects on Hexagonal Boron Nitride Properties. *Adv. Mater.* **2023**, *2306033*, 1–8. https://doi.org/10.1002/adma.202306033.
(56) Schué, L.; Stenger, I.; Fossard, F.; Loiseau, A.; Barjon, J. Characterization Methods Dedicated to Nanometer-Thick HBN Layers. *2D Mater.* **2016**, *4* (1), 015028. https://doi.org/10.1088/2053-1583/4/1/015028.
(57) Cassabois, G.; Valvin, P.; Gil, B. Hexagonal Boron Nitride Is an Indirect Bandgap Semiconductor. *Nat.*

*Photonics* **2016**, *10* (4), 262–266. https://doi.org/10.1038/nphoton.2015.277.
(58) Onodera, M.; Watanabe, K.; Isayama, M.; Arai, M.; Masubuchi, S.; Moriya, R.; Taniguchi, T.; Machida, T. Carbon-Rich Domain in Hexagonal Boron Nitride: Carrier Mobility Degradation and Anomalous Bending of the Landau Fan Diagram in Adjacent Graphene. *Nano Lett.* **2019**, *19* (10), 7282–7286. https://doi.org/10.1021/acs.nanolett.9b02879.
(59) Tsushima, E.; Tsujimura, T.; Uchino, T. Enhancement of the Deep-Level Emission and Its Chemical Origin in Hexagonal Boron Nitride. *Appl. Phys. Lett.* **2018**, *113* (3), 031903. https://doi.org/10.1063/1.5038168.

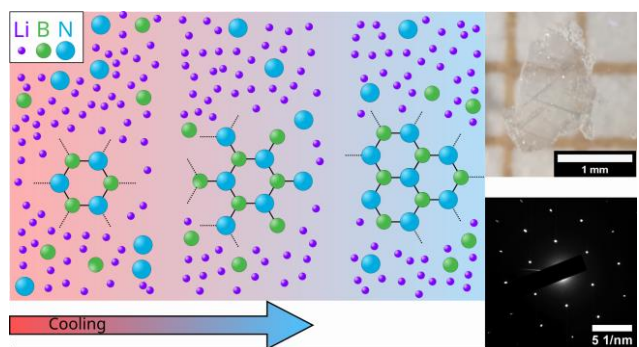

*For Table of Content only*